# Emergent Magnetic Phenomenon with Unconventional Structure in Epitaxial Manganate Thin Films

Mingwei Yang[1,2], Kuijuan Jin*[,1,2,3], Hongbao Yao[1,2], Qinghua Zhang[1], Yiru Ji[1,2], Lin Gu[1,2], Wenning Ren[1,2], Jiali Zhao[4], Jiaou Wang[4], Er-Jia Guo[1,2,3], Chen Ge[1,2], Can Wang[1,2,3], Xiulai Xu[1,2,3], Qiong Wu[5], and Guozhen Yang[1,2]

[1] Institute of Physics, Chinese Academy of Sciences, Beijing 100190, China

[2] University of Chinese Academy of Sciences, Beijing 100049, China

[3] Songshan Lake Materials Laboratory, Dongguan, Guangdong 523808, China

[4] Beijing Synchrotron Radiation Facility, Institute of High Energy Physics, Chinese Academy of Sciences, Beijing 100039, China

[5] International Center for Quantum Materials, School of Physics, Peking University, Beijing 100871, China

**Abstract**

A variety of emergent phenomena has been enabled by interface engineering in the complex oxides heterostructures. While extensive attention has been attracted to LaMnO$_3$ (LMO) thin films for observing the control of functionalities at its interface with substrate, the nature of the magnetic phases in the thin film is, however, controversial. Here, it is reported that the ferromagnetism in 2 and 5 unit cells thick LMO films epitaxially deposited on (001)-SrTiO$_3$ substrates, a ferromagnetic/ferromagnetic coupling in 8 and 10 unit cells ones, and a striking ferromagnetic/antiferromagnetic pinning effect with apparent positive exchange bias in 15 and 20 unit cells ones are observed. This novel phenomenon in both 15 and 20 unit cells films indicates a coexistence of three magnetic orderings in a single LMO film. The high-resolution scanning transmission electron microscopy suggests a P2$_1$/n to Pbnm symmetry transition from interface to surface, with the spatial stratification of MnO$_6$ octahedral morphology, corresponding to different magnetic orderings. These results should shed some new lights on manipulating the functionality of oxides by interface engineering.

**[Keywords]** Magnetism, oxide heterostructures, interface, positive exchange bias.

**Introduction**

Artificial oxide heterostructures with chemically abrupt interfaces provide a platform for engineering bonding geometries that lead to emergent phenomena.[1-8] Various interesting properties and diverse phase diagrams have been demonstrated in LaMnO$_3$ (LMO) thin films, multilayers, and superlattices, making the interfaces between LMO and substrates become an ideal candidate for discovering new phenomena for controlling functionalities.[9-12] Stoichiometric LMO bulk is known to be a layer-type (A-type) antiferromagnet, in which Mn$^{3+}$ is a Jahn-Teller ion with a $t_{2g}^3 e_g^1$ occupancy and the in-plane interaction between adjacent Mn ions is ferromagnetic (FM) while the out-of-plane one is antiferromagnetic (AFM).[13] In contrast to bulk LMO, there has been lots of controversy about magnetic phase in thin films. While some studies describe the appearance of FM behavior in stoichiometric thin films,[9,14,15] other reports have shown that antiferromagnetic order staying in the films thinner than 5 unit cells (u.c.), and ferromagnetic phase in films thicker than 5 u.c..[9,13] Meanwhile, almost all experiments have unveiled an *insulating* feature[9,12,13,16,17] for LMO films in both ferromagnetic and antiferromagnetic phases, although many theoretical reports have predicted that LMO thin films possess a Pbnm structure and with a *ferromagnetic metallic* phase.[18,19] Numerous efforts have been devoted to eliminate this paradox by reducing sample symmetry[15] or confirming electronic phase separation,[14] among which H. J. Xiang *et. al*[15] firstly introduced unconventional P2$_1$/n structure in LMO to explain this *ferromagnetic insulating* feature. However, this novel P2$_1$/n structure in LMO films has not been experimentally observed so far.

In this paper, we firstly report an experimental observation of *ferromagnetic insulating* state in ultra-thin LMO films with thickness of 2 and 5 u.c.. We demonstrate a ferromagnetic/ferromagnetic (FM/FM) coupling in 8 and 10 u.c. films, indicating the appearance of a harder FM state in the region above 5 u.c.. In addition to the FM/FM coupling, a striking ferromagnetic/antiferromagnetic (FM/AFM) pinning effect with remarkable positive exchange bias is shown in 15 and 20 u.c. thin films, indicating an antiferromagnetic order appearing in the layer above 10 u.c. away from the interface. To reveal the evolution of magnetic property in different regions, we show the characterizations by high-resolution scanning transmission electron microscopy (STEM), which display a transition from $P2_1/n$ to Pbnm structure, corresponding to the evolution of FM to AFM. Our study demonstrates that different magnetic orderings, diverse octahedral morphologies, and various exchange couplings can be achieved in a single oxide by controlling their dimensionality.

**Results and Discussion**

The crystalline quality of LMO films with different thicknesses is ensured by high-resolution x-ray diffraction (XRD) patterns in Figure 1a, where only peaks from substrates and films could be observed. Reciprocal space mapping (RSM) results reveal the coherent growth of samples with thickness up to 20 u.c., while the 200 u.c.-thick film exhibits a slight lattice relaxation in *ab* plane. Besides, all samples present lattice relaxation along *c* direction (Figures 1b-f).

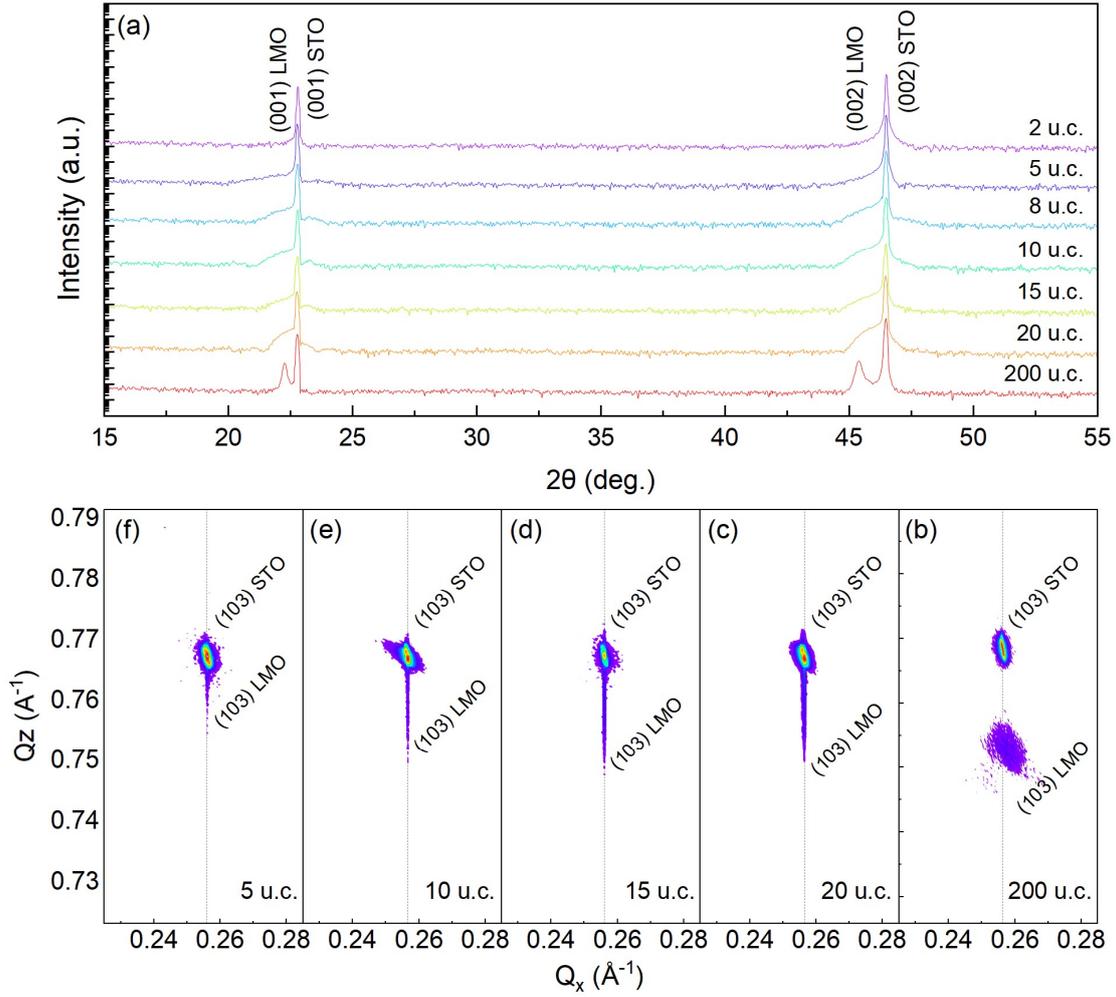

**Figure 1.** a) X-ray diffraction (XRD) results of LaMnO$_3$ (LMO) films with thickness of 2, 5, 8, 10, 15, 20, and 200 u.c.. The data have been offset for better view. b-f) Reciprocal space mapping (RSM) of LMO with thickness of 5, 10, 15, 20, and 200 u.c., respectively.

Figure 2 shows the in-plane magnetic hysteresis (MH) loops of LMO films with various thicknesses at 10 K. The ferromagnetic nature in 2 u.c.-thick LMO films is confirmed by MH loop (Figure 2a). This ferromagnetism in 2 u.c. films is firstly obtained, whereas it was reported as an antiferromagnetic layer by other groups.[9,13,14] This ferromagnetic feature has also been found in 5 u.c. samples displayed in Figure 2b. With increasing film thickness up to 8 or 10 u.c., interestingly, the MH loops become step-like (Figures 2c-d) and are composed by two regular MH loops with different coercive field and saturation magnetization ($M$s). This superimposition indicates that two different FM layers present in these films. For the simplicity to be mentioned later, we name it as the first ferromagnetic layer (FML1) with thickness roughly goes from 1

- 6 u.c.. The second ferromagnetic layer (FML2) means the layer with harder ferromagnetic feature (with larger coercive filed) approximately from the 7 - 11 u.c.. Besides, the FM/FM interaction between FML1 and FML2 brings a distinct exchange spring effect in these films. This step-like feature retains in 15 (Figure 2e), 20 (Figure 2f) and 200 u.c. (Figure 2g) films, suggesting the stable stratification and reproducibility of these two FM layers. From Figure 2e, it can be seen that a striking positive exchange bias of ΔH = + 931 Oe in 15 u.c. LMO films occurs, and the positive exchange bias remains appreciable in 20 u.c. films (Figure 2f). The appearance of positive exchange bias indicates the coexistence of a FM layer, an AFM layer, and a FM/AFM coupling between these two layers, implying an appearance of AFM phase in LMO films[10,20,21] following the FML2 in the films. To be mentioned later, we denote the layer thicker than 11 u.c. as the antiferromagnetic layer (AFML). The existence of this antiferromagnetic order is also verified by the abrupt drops in the MH loops around zero magnetic field, indicated by red circles (Figure 2g). These drops correspond to the spin flip of the AFML when subjected to a certain external magnetic field.[21-23] This antiferromagnetic feature is analogous to the antiferromagnetic property of LMO bulk.[24,25] Besides, the slight increase of magnetic moment ($M$) in 15 (200) u.c. films from that in 10 (15 and 20) u.c. films shown in Figure 2h, suggesting an antiferromagnetic order in LMO films thicker than 10 u.c.. The corresponding spin orderings are schematically shown in Figures 2i-k.

From above results, we can deduce that FML1, FML2, and AFML emerged successively in a single LMO film with the thickness of 15, 20, or 200 u.c. along with sample growing. This emergent magnetic phenomenon is, to our knowledge, unprecedentedly reported.

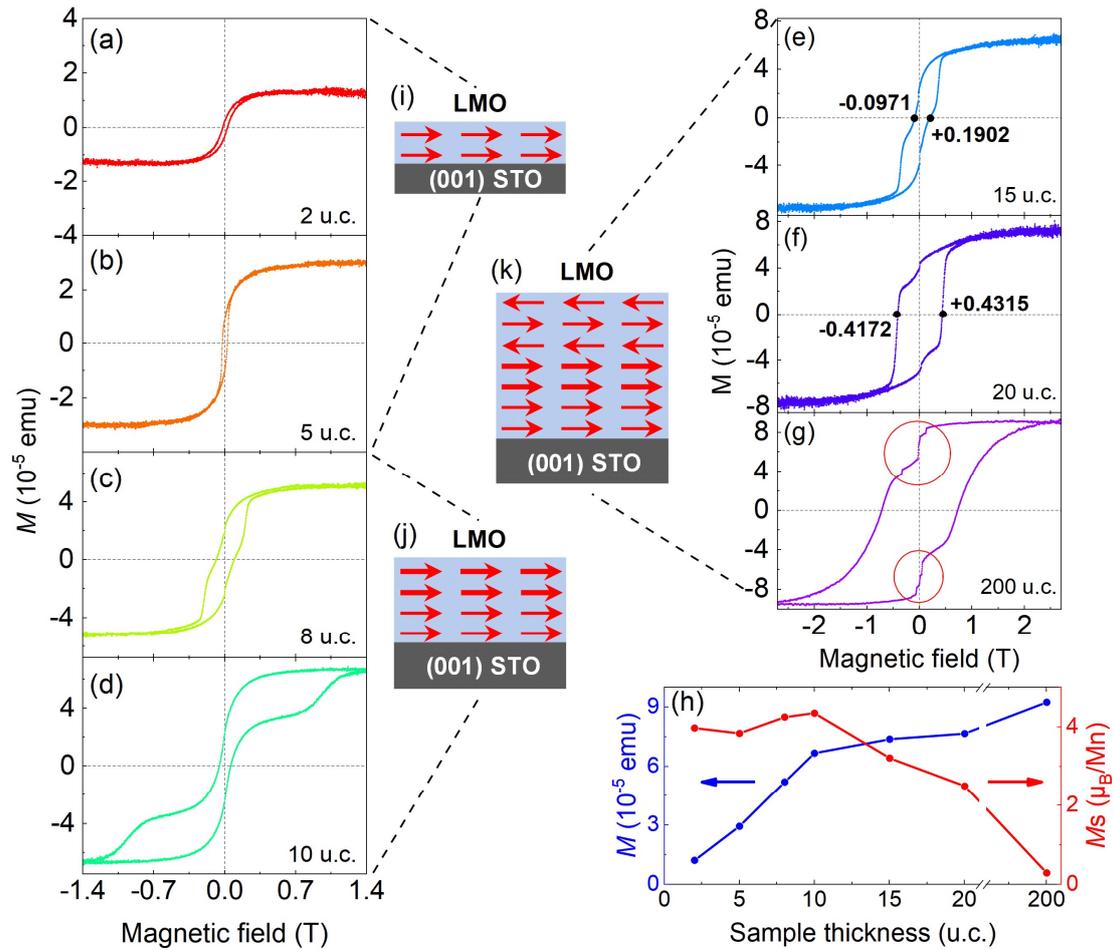

**Figure 2.** a-g) In-plane magnetic hysteresis (MH) loops for LaMnO$_3$ (LMO) films with the thickness of 2, 5, 8, 10, 15, 20, and 200 u.c. at 10 K, respectively. The numbers in (e) and (f) indicate the corresponding coercive fields pointed out by the black dots. h) The absolute magnetic moment (*M*) and saturation magnetization (*M*s) of LMO films as a function of thickness. The dots are experimental data and the lines are for eye guide. i-k) Schematic illustration of the spatial distribution of spin in LMO films with thickness of 2 and 5 u.c., 8 and 10 u.c., and 15, 20, and 200 u.c., respectively. The direction of arrows indicates the spin orientation, and their widths represent the relative strength of spin.

Figures S1a-c illustrate the x-ray absorption spectroscopy (XAS) for La (*M*-edge), Mn (*L*-edge), and O (*K*-edge), respectively, of LMO films with different thicknesses. A direct comparison from the electron energy loss spectroscopy (EELS) spectra collected for La, Mn and O, respectively, of a 15 u.c. LMO film is shown in Figures S1d-1f. The continuous blue shift of the $L_3$ edge of Mn from interface to surface (Figure S1f) suggests that Mn$^{2+}$ is locally confined in LMO near the interface in all samples regardless of their thickness, and Mn$^{3+}$ dominates subsequent area. This spatial distribution of Mn$^{2+}$ and Mn$^{3+}$ is consistent with the fading of Mn$^{2+}$ characteristic peak

at 640.4 eV in XAS (Figure S1b) with increasing thickness where the ratio of $Mn^{2+}/Mn^{3+}$ becomes smaller. This distribution is also in good agreement with the enhancement of the O 1s peak at 529.8 eV with increasing thickness (Figure S1c) which represents higher Mn oxidation states than +2.[26] The appearance of $Mn^{2+}$ may attribute to the electronic reconstruction at the interface between STO and LMO. It is known that the net charge imbalance between two adjacent sublayers will inevitably lead to the polar catastrophe. Therefore, the charge transfer between two transition metal ions happens to neutralize the charged interfaces.[17,27] For an ultrathin LMO film with a thickness below 2 u.c., we believe that surface effect may also contribute to add free electrons, e.g. oxygen vacancies or strain-induced nonstoichiometry, into LMO film, leading the partial $Mn^{3+}$ ions change to $Mn^{2+}$ ions. [28,29,30,31]

To reveal the mechanism of our magnetic results, the unit-cell-resolved structural evolution is characterized with high-resolution scanning transmission electron microscopy (STEM). Figure 3a shows the inverted annular bright-field (ABF) cross section of a 15 u.c. film with a clear interface indicated by the white dashed line. The uniform atomic distribution presented here shows a good quality of the sample. A large-area STEM result is also illustrated in Figure 3e. Referring to STEM imaging, FML1, FML2, and AFML are marked in Figure 3b, and the corresponding oxygen octahedra are schematically illustrated in Figure 3c using Visualization for Electronic and Structural Analysis (VESTA) program,[31] where the octahedra manifest vast diversity from each layer. We carried out statistics over 50 atomic columns and 22 atomic rows from the STEM image shown in Figure 3e using DigitalMicrograph. The statistical example for a single row in FML1, FML2, and AFML is shown in Figure S2. For the substrate, generally, the cubic STO presents virginal $TiO_6$ octahedra and belongs to typical $Pm\bar{3}m$ space group. In both FML1 and FML2, two kinds of octahedra, denoted as OC1 and OC2, appear alternatively in G type manner[15] with three-dimensional alternation of OC1 and OC2. Figure 3d shows O-O length ($L_{O-O}$) in *ab* plane ($L$x) and *c* axis ($L$z) of the octahedra denoted as OC2 for one row and OC1 for the nearest rows. $L$z ($L$x) is always longer (shorter) in OC1 than the one in OC2, shown as the zigzag curve in Figure 3d for both FML1 and FML2. In Figure 3d, we find that both OC1 and OC2 are *c*-axially elongated and $L$z is longer than $L$x for FML1 due to the compressing effect from the substrate. This characteristic is verified by the statistics of STEM characterization for 5 u.c. LMO films displayed in Figure S3. However, for FML2, $L$z is longer (shorter) than $L$x in OC1 (OC2). The alternation of OC1 and OC2 in FML1 matches perfectly with the octahedral spatial arrangement of LMO in the P2$_1$/n structure predicted by H. J. Xiang *et. al.*.[15] Further evidence on the octahedra tilt is illustrated by systematic analysis of Mn-O-Mn bond angle ($\theta_{Mn-O-Mn}$) across the film growth direction (Figure S4). The FML1 possesses strongly tilted octahedra and $\theta_{Mn-O-Mn}$ changes to ~ 172º after the first 2 u.c.. The octahedral tilt, i.e. $\theta_{Mn-O-Mn}$, keeps almost constant value with standard error bar in both FML1 and FML2. In AFML, the $\theta_{Mn-O-Mn}$ decreases slightly to 170º. The changes in $\theta_{Mn-O-Mn}$ agree well with the significant trend in the O-O length shown in Figure 3d, indicating a robust modulation of MnO$_6$ octahedra as increasing LMO film thickness. Please note that our STEM measurements had been performed over at least five different locations on the same sample. All results suggest that the thickness-driven octahedral tilt within LMO layer is uniformly distributed parallel to the STO substrate. In addition, the insulating behaviors measured for our films shown in Figure S5 is in good agreement with the insulating phase for ferromagnetic P2$_1$/n structure. The gradual structural transition in LMO films can be attributed to the symmetry-mismatch between STO (cubic) and LMO (orthorhombic).[9,13] The relaxation of shear strain will inevitably change the crystallographic symmetry of LMO films with increasing thickness. This substrate/film lattice and symmetry mismatch is the main source of the octahedral rotation modification and Jahn-Teller distortion.[12,14,32-34] The evolution of MnO$_6$ octahedra changes bond length and angle, thus affecting the balance between the intra-atomic exchange interaction energy and crystal field energy through the structural distortion. As a result, the electronic ground state is modified, leading to the observed emerging magnetic phenomenon in our LMO films.[34-37]

In AFML, $L$x ($L$z) remains roughly the same for OC1 and OC2, with $L$z being longer than *L*x. The similarity of OC1 and OC2 in AFML with *c*-axially stretching and *ab* plane compressing remains, which is described as normal $Q_3$ distortion mode.[25,38] The octahedra in AFML manifest the heaviest rotation (Figure S4) and the film demonstrates the Pbnm structure.

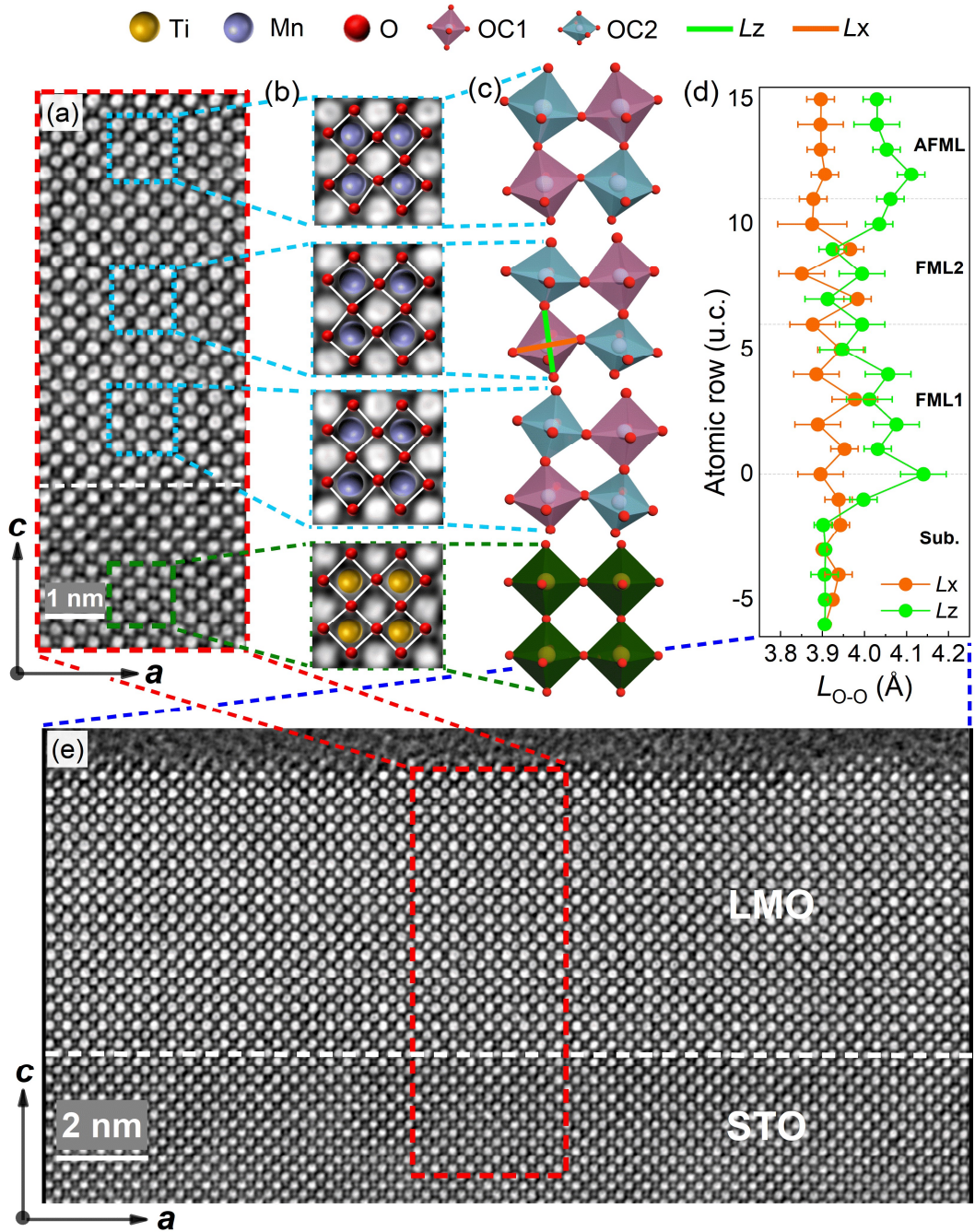

**Figure 3.** a) Inverted annular bright-field scanning transmission electron microscopy (ABF-STEM) image of a 15 u.c. LaMnO$_3$ (LMO) film from (e). The white dashed line in (a) represents the interface. b) Illustration of octahedra in dashed boxes referring to (a) in substrate (sub.), first ferromagnetic layer (FML1), second ferromagnetic layer (FML2), and antiferromagnetic layer (AFML). c) Enlarged schematic diagrams of octahedra corresponding to (b). The brown,

mediumslateblue, and red disks denote Ti, Mn, and O atoms, respectively. The purple and dark cyan octahedra, denoted as OC1 and OC2, represent octahedra in two nonequivalent positions, respectively. d) Evaluated O-O length ($L_{O-O}$) in *ab* plane ($L$x) and *c* axis ($L$z) of octahedra in each layer. The dashed lines represent layer boundaries. e) A large-area cross section of the 15 u.c. LMO film from scanning transmission electron microscopy. The white dashed line in (e) represents the interface.

Above relative variations of $L$x and $L$z in FML1, FML2, and AFML refer to the ferromagnetic, harder ferromagnetic, and antiferromagnetic phases in the films, respectively. The octahedra are schematically plotted in Figure 4 for P2$_1$/n (Figure 4a) and Pbnm (Figure 4b), respectively, as well as the occupation of Mn *3d* orbitals (Figures 4c-d). In the alternation of OC1 and OC2, the single $e_g$ electron of OC1 occupies the $d_{3z^2-r^2}$ orbital with lower energy, and the $e_g$ electron of OC2, however, occupies the $d_{x^2-y^2}$ orbital as shown in Figures 4a and 4c. Three-dimensionally alternating $d_{3z^2-r^2}/d_{x^2-y^2}$ orbital order gives rise to the ferromagnetism in FML1 and FML2 according to Goodenough-Kanamori rules. Detail explanations can be found in Reference [15]. Since $L$z is always larger than $L$x in the same octahedron in FML1, and their relative length alternates three-dimensionally in FML2 (Figure 3d), the orbital overlaps of half-filled $d_{3z^2-r^2}$ in OC1 and empty $d_{3z^2-r^2}$ in OC2 with O$_{2p}$ are enlarged in FML2 comparing to those in FML1, bringing more distinct MH loops with a larger coercive field and *M*s in FML2 than that of FML1, as shown in Figures 2c-d. In AFML, as shown in Figures 4b and 4d, the hybrid orbital $e_{g1}$ Mn ion interacts with the empty $e_{g2}$ orbital of adjacent Mn through O$_{2p}$ orbital with a negative exchange integral along *c* axis, resulting in an antiferromagnetic super-exchange interaction.[25,39,40] As a result, this layer shows an A-type antiferromagnetic feature like the bulk. We note that the emergent magnetic phenomenon in this work has not been reported by other groups. We believe that the correct stoichiometry and negligible oxygen vacancy present in our LMO film may be one of the key factors that control the magnetic ground state.

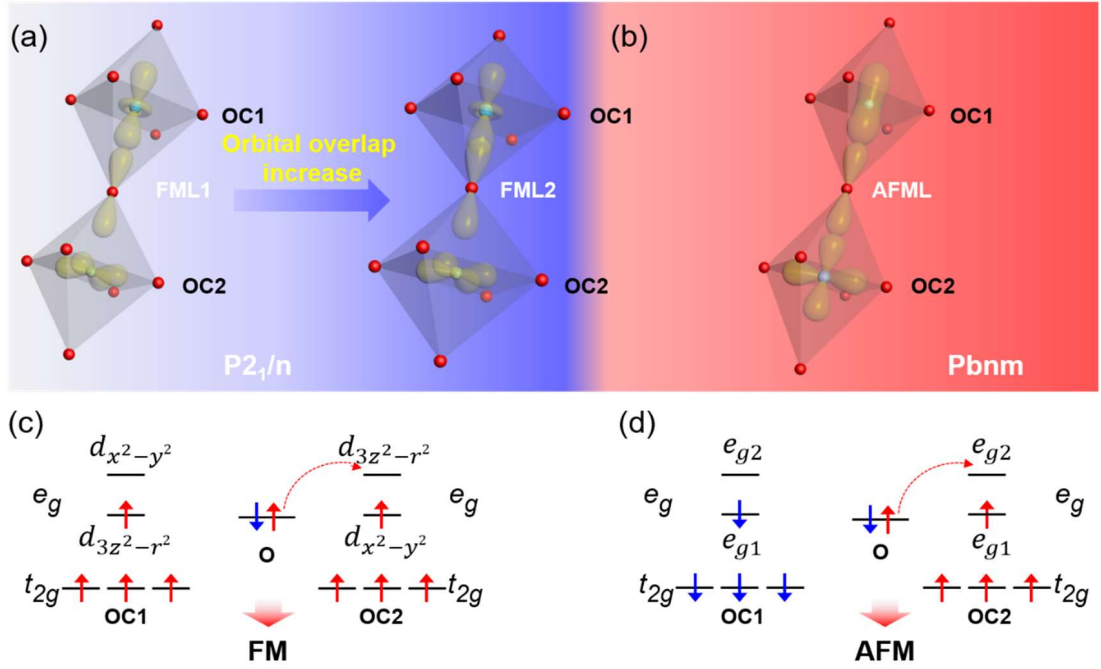

**Figure 4.** Schematic illustration of orbital and spin configuration of Mn ions in the first ferromagnetic layer (FML1), second ferromagnetic layer (FML2), and antiferromagnetic layer (AFML). a) The orbital overlap in FML1 and FML2 that exhibit P2$_1$/n feature. b) Orbital overlap in AFML that manifests Pbnm characteristic. c,d) Orbital occupation and spin configuration of Mn ions in OC1 and OC2 in LaMnO$_3$ (LMO) with P2$_1$/n and Pbnm structure, respectively. The red and blue arrows indicate spin-up and spin-down state, respectively. The $e_{g1}$ and $e_{g2}$ represent hybridization levels.

**Conclusion**

We studied the evolution of the insulating magnetic LMO thin films and found that three magnetic layers with FM, harder FM, and AFM features arising successively with increasing film thickness. A FM/FM coupling in 8 and 10 u.c. films, and a striking FM/AFM pinning effect with remarkable positive exchange bias in 15, 20, and even up to 200 u.c. films are demonstrated. The statistics for high-resolution STEM characterization displays a transition from P2$_1$/n to Pbnm structure, revealing the mechanism of the magnetic property evolution from FM to AFM for the films. Our results also demonstrate that various magnetic orderings, diverse octahedral morphologies, and different exchange couplings can be achieved in a single oxide by controlling the interfacial effect, paving a route to manipulate the functionalities in heterostructure through interface engineering.

**Experimental Section**

*Sample synthesis*: The LMO films of various thickness were deposited on (001)-oriented STO substrates at 680 °C using a pulsed laser deposition (PLD) system. During the deposition, the target-substrate distance was set to of 7.2 cm and the oxygen partial pressure was adjusted to 0.1 Pa. A XeCl laser with central wavelength of 308 nm was employed to provide a laser energy density of 1.2 J/cm$^2$ at a frequency repetition of 3 Hz. After the growth, the LMO films were annealed *in situ* for 7.5 min in a vacuum of ~2 × 10$^{-3}$ Pa to eliminate excess oxygen, and then were cooled down to room temperature at a rate of 15 °C/min.

*X-ray characterization*: The XRD and RSM analyses were performed using a Rigaku SmartLab (8 kW) high-resolution x-ray diffractometer, with the wavelength of the x-ray is 0.154 nm. The x-ray absorption spectroscopy (XAS) was collected under the total electron yield (TEY) mode at the 4B9B line station of Beijing Synchrotron Radiation Facility.

*Physical properties analyses*: The magnetic properties of these samples were measured with a vibrating sample magnetometer (VSM) of physical properties measurement system (PPMS). The external magnetic field is applied from -3 T to 3 T parallel to *ab* plane when measuring MH loops at 10 K. The resistivity was measured using van der Pauw methods with PPMS.

*Scanning transmission electron microscopy*: The atomic structures of these heterostructures were characterized using an ARM-200CF transmission electron microscope operated at 200 keV and equipped with double spherical aberration (Cs) correctors.

*Electron energy loss spectroscopy*: The unit-cell-resolved EELS spectra were collected using STEM at the same time with structural characterization of these films.

*Statistical analysis*: We imported the STEM images in Figure 3e and Figure S3a into DigitalMicrograph to obtain the coordinates of each atoms. Using their coordinates, we calculated the O-O length ($L$x and $L$z) in each octahedron and the M-O-M angle ($\theta_{\text{M-O-M}}$) between adjacent octahedra. As the LMO sample shows alternation of OC1 and OC2, the O-O length and θ$_{M-O-M}$ is collected every other column. The dots in Figure 3d, Figure S3b, and Figure S4b represent the mean value over odd atomic column along thickness direction, and the error bar comes from the standard deviation. Figure S3 shows the statistics of O-O length of the atomic row in substrate, FML1, FML2, and AFML, indicated by black arrows, along *a* axis. The four left (right) bar charts represent the O-O length on odd (even) atomic columns.

**Acknowledgments**

This work was supported by the National Key Basic Research Program of China (Grant 2019YFA0308500 and 2020YFA0309100), the National Natural Science Foundation of China (Grant Nos. 11721404, 51761145104, 11974390 and 11674385), the Key Research Program of Frontier Sciences of the Chinese Academy of Sciences (Grant No. QYZDJ-SSW-SLH020), and the Youth Innovation Promotion Association of Chinese Academy of Sciences (Grant No. 2018008).